\documentclass[aip]{revtex4-1}
\usepackage{graphicx}
\usepackage[version=4,arrows=pgf-filled,
textfontname=sffamily,
mathfontname=mathsf]{mhchem}
\usepackage{bm}
\usepackage{fancyhdr}
\usepackage{amsmath}
\usepackage{amsfonts}
\usepackage{amsbsy}
\usepackage{amssymb}
\usepackage[mathscr]{eucal}
\usepackage{color}
\usepackage{soul}
\usepackage{mathtools}
\usepackage[english]{babel}
\let\originalselectlanguage\selectlanguage
\renewcommand{\selectlanguage}[1]{%
  \def\langinput{#1}%
  \ifnum\pdfstrcmp{\langinput}{en}=0
    \originalselectlanguage{english}%
  \else\ifnum\pdfstrcmp{\langinput}{eng}=0
    \originalselectlanguage{english}%
  \else
    \originalselectlanguage{#1}%
  \fi\fi
}

\begin{document}
\title{Modern view of activated rate processes: unidirectional fluxes at equilibrium, correlation functions, and splitting probabilities.}

\author{Alexander M. Berezhkovskii}
\email{tatyber2002@yahoo.com}
\affiliation{Bethesda, MD, 20814}
\author{Kevin Song}
\email{kcsong@utexas.edu}
\affiliation{Oden Institute for Computational Engineering and Sciences, University of Texas at Austin, Austin, TX, 78712}
\author{Dmitrii E. Makarov}
\email{makarov@cm.utexas.edu}
\affiliation{Oden Institute for Computational Engineering and Sciences, University of Texas at Austin, Austin, TX, 78712}
\affiliation{Department of Chemistry, University of Texas at Austin, Austin, TX, 78712 }

\begin{abstract}
More than 80 years ago Kramers published a paper calculating how fast a Brownian particle escapes  from a potential well over an activation barrier. Since then Kramers' model has been widely adopted by nuclear physics, biophysics and chemical physics communities as a description of activated barrier crossing. From a chemical kinetics perspective, Kramers' theory provides a mapping from continuous dynamics to discrete-state chemical kinetics. Motivated by recent developments, this  Perspective provides a rigorous way of performing such a mapping, explaining why and how Kramers' theory works from several points of view. Specifically, we consider transitions of a Brownian particle between two potential wells corresponding to the ``reactant'' and the ``product'' of a chemical reaction.  A central unifying idea is to divide the equilibrium ensemble of possible states of the system into two sub-ensembles corresponding to the reactant and product states and then to consider fluxes  between these sub-ensembles. Importantly, naive separation based on the location measured relative to the barrier top does not result in a mapping that is physically tenable, and instead the past of the trajectory should be considered. Thus constructed reactant and product ensembles provide an internally consistent description of the problem when also viewed from two different perspectives: one based on the definition of the rate as a conditional transition probability per unit time and the other based on the relaxation modes of the time-evolution operator governing the dynamics.     
\end{abstract}

\maketitle

\section{Introduction}

Chemical kinetics provides a discrete -- and thus coarse-grained -- view of molecular dynamics. Without this view, not much progress would have been made interpreting chemical experiments. Beyond chemistry, this approach proved to be productive in numerous applications arising in biophysics\cite{hill_free_1989,phillips_physical_2013,kolomeisky_motor_2015,makarov_single_2015}, stochastic thermodynamics\cite{seifert_stochastic_2025,peliti_stochastic_2021}, and mechanics at microscopic or mesoscopic scales\cite{zhao_resistance_2011,makarov_kinetic_2001,seifert_rupture_2000}. When a chemist thinks of, say, chair-boat isomerization, or a biochemist considers protein folding, they effectively lump a continuum of molecular structures around the equilibrium configurations or all possible structures of a polypeptide chain into a pair of states, A and B,  called ``reactant'' and ``product'' (which could be the chair/boat isomers or the folded/unfolded proteins).  This results in a kinetic scheme of the form
\begin{equation} \label{AB}
\ce{A <=>  B}
\end{equation}
If we denote the fractions of A and B in the system $P_A(t)$ and $P_B(t)$, then phenomenological chemical kinetics predicts those to evolve in time according to the rate equations  
\begin{equation} \label{masterEq}
    dP_A/dt=-dP_B/dt = -k_{A\to B}P_A(t)+k_{B\to A}P_B(t), 
\end{equation}
 where $k_{A\to B}$ and $k_{B\to A}$ are the phenomenological rate coefficients for the transitions from A to B and from B to A.  These equations can also be used to describe the same system at a single-molecule level, as observed in molecular simulations or in single-molecule studies. In this case $P_A$ and $P_B$ are to be interpreted as the probabilities of finding the molecule in the states A and B. In equilibrium, the molecule will switch stochastically between A and B. The equilibrium probabilities $P_A$ and $P_B$ can be understood as the fractions of time the molecule spends in A and B, and they satisfy the detailed balance relationship
 \begin{equation} \label{detailed balance}
     k_{A\to B}P_A = k_{B\to A}P_B\equiv J_{A\to B}=J_{B\to A},
 \end{equation}
defining the {\it unidirectional fluxes} from A to B ($J_{A\to B}$) and from B to A ($J_{B\to A}$) at equilibrium. These fluxes can be understood as the numbers of transitions,  per unit time, that each molecule makes in either direction. Eq.~\ref{detailed balance} expresses the dynamical nature of chemical equilibrium, which results from the balance between the transitions in the opposite directions.

How can we map the dynamics of the molecule in phase space onto such a phenomenological description? One can partition the entire phase- or conformational space of a molecule into two domains  declared to be A and B, an idea that goes back to the work of Marcelin\cite{marcelin_notitle_1915}, Wigner\cite{wigner_notitle_1932} and Eyring\cite{eyring_activated_1935}. Unfortunately, this way of mapping (which leads to what is known as transition state theory -- see, e.g., refs.\cite{peters_reaction_2017,pollak_reaction_2005,elber_molecular_2020} for a comprehensive review and discussion) typically results in microscopic unidirectional fluxes (crossing the boundary between the two domains) that are greater than the phenomenological fluxes observed experimentally -- a phenomenon known as the ``recrossing problem'', because quick recrossings of the surface dividing the A and B domains should not be counted as true transitions\cite{shui_monte_1972,chandler_notitle_1978,bennett_molecular_1977}.  

A closely related to this technical issue is a practical one concerning evaluation of $k_{A\to B}$. This rate coefficient is usually described by the Arrhenius law:
\begin{equation} \label{Arrhenius}
    k_{A\to B}=\nu e^{-E_a/k_B T},
\end{equation}
where $E_a$ is the activation energy and $\nu$ is a prefactor often interpreted as an ``attempt frequency'' or a ``speed limit'' of the reaction. Transition state theory predicts the latter to be comparable with molecular vibrational frequencies  (say in an inverse picosecond range \cite{elber_molecular_2020}),   which is a reasonable approximation for gas-phase chemical reactions. But when the Arrhenius law is invoked to describe the rate of condensed-phase reactions in solution, and in particular of biomolecular processes such as protein folding and unfolding\cite{klimov_viscosity_1997,socci_diffusive_1996}, the value of the prefactor turns out to be many orders of magnitude lower\cite{w_a_eaton_fast_2000,li_theoretical_2003,kubelka_protein_2004}.   The model introduced by Kramers  in 1940 offers a solution to both problems. Kramers proposed to use theory of Brownian motion as an intermediate-level (i.e., not fully microscopic) description of chemical and nuclear reactions\cite{kramers_brownian_1940}. In Kramers' view one special molecular degree of freedom $x$ (e.g., a molecular vibration) is treated as what we now call ``the reaction coordinate'', while the effect of the remaining degrees of freedom including those of the solvent in solution-phase reactions can be described as leading to a friction force proportional to velocity  as well as to a random force resulting from random molecular impacts. In other words, Kramers proposed to treat the motion along $x$ as that of a Brownian particle subjected to some effective potential $U(x)$.  This can be described by the Langevin equation (see, e.g., refs.\cite{robert_zwanzig_nonequilibrium_2001,nitzan_chemical_2013})
\begin{equation} \label{Langevin}
    m\frac{d^2 x}{dt^2}=-\frac{dU}{dx}-\gamma \frac{dx}{dt}+\xi(t), 
\end{equation}
where $m$ is the particle's mass, $\gamma$ is a friction coefficient, and $\xi(t)$ is a Gaussian, delta-correlated random force with zero mean that  satisfies the fluctuation-dissipation theorem, 
\begin{equation} \label{FDT}
    \langle \xi(t) \xi(t') \rangle = 2 k_B T \gamma \delta(t-t'). 
\end{equation}
In the absence of the potential, the dynamics of a particle at sufficiently long timescale is diffusive\cite{robert_zwanzig_nonequilibrium_2001}, and the friction coefficient is related to the particle's diffusivity $D$ by the Stokes-Einstein relationship
\begin{equation} \label{stokes einstein}
    D=\frac{k_B T}{\gamma}.
\end{equation}

For moderate values of the friction coefficient Kramers' theory recovers the transition-state-theory result given by the Arrhenius law (Eq.\ref{Arrhenius}) with $\nu \sim \omega_A/2\pi$, where $\omega_A$ is the vibrational frequency along $x$  near the minimum of $U(x)$.  But for sufficiently large friction, one finds that the escape rate is inversely proportional to the friction coefficient $\gamma$ thus explaining why, in solution, the prefactor $\nu$ could be orders of magnitude lower than that predicted by transition state theory. Moreover, later work \cite{tannor_derivation_1994,pollak_theory_1986} showed that Kramers' estimate for $k_{A\to B}$ solves the recrossings problem. Although Kramers  considered escape from a potential well over a barrier, which  -- effectively -- can be thought of as a model of an irreversible chemical reaction $\ce{A ->  B}$, his calculation is readily extended to include a second potential well (B) and consider the reverse reaction of escaping this well in the opposite direction. When this is done, Kramers' theory then provides a proper mapping between continuous Langevin dynamics, Eq.~\ref{Langevin} and chemical kinetics, Eq.~\ref{masterEq}.  How does Kramers' theory achieve this? 

In view of the fact that Kramers' theory is widely adopted by multiple communities including those who study chemical dynamics, nucleation, nuclear reactions, biomolecular folding, and other disciplines, we feel it would be beneficial to discuss the precise assumptions behind Kramers' theory, consider it from several different perspectives, and explain why it describes the dynamics of experimental observables as well as it does.  A comprehensive review\cite{hanggi_50_1990} on this subject was written to celebrate the 50-th anniversary of Kramers' paper. The goal of the present Perspective is to discuss Kramers' results in light of the insights that have emerged over the last 3 decades. Those include  new understanding of what a ``good reaction coordinate'' means\cite{du_transition_1998,e_transition-path_2010,peters_reaction_2013,peters_reaction_2017, berezhkovskii_one-dimensional_2005,berezhkovskii_diffusion_2013}, ideas from transition path theory\cite{vanden-eijnden_exact_2009,bolhuis_transition_2002,hummer_transition_2004,warmflash_umbrella_2007,roux_transition_2022}, a multitude of computational  ``celling'' techniques\cite{elber_molecular_2020,elber_perspective_2016} (such as Markov state models and milestoning) that aim at mapping continuous dynamics onto discrete processes, and  the development of single-molecule experiments that have made observations of stochastic dynamics of chemical reactions under equilibrium conditions the domain of an experimentalist (see, e.g., refs.\cite{chung_protein_2018,schuler_protein_2008,hoffer_probing_2019,schuler_single-molecule_2013,makarov_single_2015,haran_single-molecule_2026}). At the same time, this work is not meant to be a comprehensive review; rather there is a single central question that we wish to address here. The question is: how does one map continuous dynamics that is the starting point of  Kramers'  model onto a discrete-space continuous-time Markov-jump process described by the equations of chemical kinetics? In other words, how does one assign a ``microscopic'' configuration of a Langevin particle, specified by its position and velocity, to belong to the reactant or to the product state? In what follows, we first discuss this assignment, building on the developments from the last 25 years that bring out the importance of transition paths\cite{hummer_transition_2004}. We show that a proper classification of microscopic states as reactants and products, while in a sense inconsistent with equilibrium thermodynamics, allows one to derive Kramers' results for strong and moderate friction by calculating the unidirectional flux of either reactants or products into the other state at equilibrium. This approach is rather general  and applicable to any potential barrier (or even a potential with no barrier), not just a quadratic one assumed by Kramers. We then explore the connection of this approach to two other methods of calculating the reaction rate -- one based on correlation functions (as originally formulated by Bennett\cite{bennett_molecular_1977}, and Chandler\cite{chandler_notitle_1978}) and the other based on analyzing the slowest relaxation mode of the system's Fokker-Planck equation\cite{berezhkovskii_diffusive_2021,roux_transition_2022,melnikov_theory_1986} -- and show that all of these methods predict the same result for the reaction rate, which is identical to that derived by Kramers.        

\section{Mapping diffusive dynamics onto two-state rate equations}

\begin{figure}
    \centering
    \includegraphics[width=0.75\linewidth]{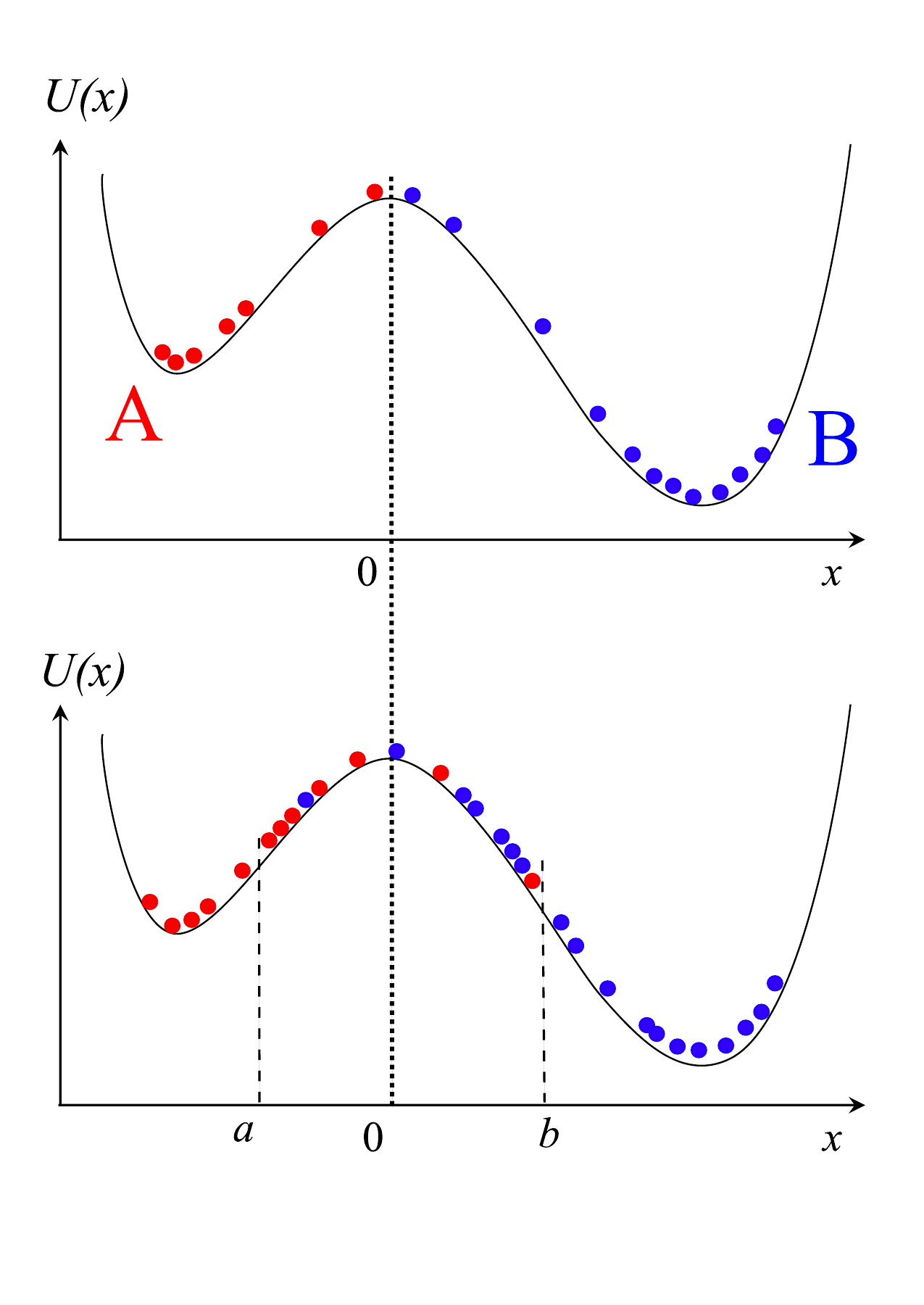}
    \caption{Mapping of continuous dynamics in a potential of mean force $U(x)$ onto two states. Each point represents the position $x$ of a realization of the system within the equilibrium ensemble and is assigned a discrete state A (red) or B (blue). Top: each point to the left of the barrier (located at $x=0$) is in state A and each point to the right is in state B. Bottom: In the transition region $(a,b)$, each trajectory that last hit the boundary $a(b)$ rather than $b(a)$ is assigned to be in A(B). To the left of $a$ the system is always in state A and to the right of $b$ in state B.}
    \label{fig:mapping}
\end{figure}

Consider a bistable system described by a ``reaction coordinate'' $x$. In equilibrium,  the system's thermodynamic properties are determined by the potential of mean force, or free energy $U(x)$, which is related to the equilibrium distribution of $x$, $p_{eq,x}(x)$, by 
\begin{equation} \label{PMF}
    p_{eq,x}(x) = \frac{1}{q}e^{-\beta U(x)},
\end{equation}
where  $\beta = \frac{1}{k_B T}$ is the inverse thermal energy, and the quantity 
\begin{equation} \label{partition function}
q=\int_{-\infty}^{\infty}dx e^{-\beta U(x)} 
\end{equation}
ensures normalization. Throughout this paper, it will be assumed that $U(x)$ has a shape shown in Fig. \ref{fig:mapping}, with two potential wells  separated by a barrier that is much greater than the thermal energy $k_B T$.  We wish to infer from this picture a chemical kinetics view, with two chemical species interconverting according to the first-order chemical kinetics scheme, Eq.~\ref{AB}. That is, we wish to map the dynamics along $x$ onto equations of first-order chemical kinetics in the form of Eq.~\ref{masterEq}.

Given the assumption of a high barrier, most of the probability density $p_{eq}(x)$ is concentrated in the vicinities of the two minima, and the most intuitive way of assigning the conformations $x$ to two states is to draw a dividing line at some point near the barrier (here, we choose it to be the top of the barrier, which, without loss of generality, will be located at $x=0$) and to declare that $x<0$ corresponds to A and $x\ge0$ to B (Fig.~\ref{fig:mapping}, top). Thus we write, for the equilibrium populations of A and B, 
\begin{equation} \label{equilPop}
    P_A=1-P_B =\frac{\int_{-\infty}^{0} dx e^{-\beta U(x)}}{\int_{-\infty}^{\infty} dx e^{-\beta U(x)}}.
\end{equation}  
The precise upper integration limit in the numerator of Eq.~\ref{equilPop} is unimportant, since the integrand is concentrated near the potential minima and is negligible near the barrier top. From the thermodynamics perspective, therefore, this is a perfectly reasonable mapping from the continuous description to a discrete one. 

We now turn to the rate coefficients. In equilibrium, Eq.~\ref{masterEq} reads 
\begin{equation}
    0=-J_{A\to B}+J_{B\to A} \equiv -k_{A\to B}P_A+k_{B\to A}P_B,
\end{equation}
where $J_{A\to B}$ is the equilibrium flux from A to B, which is compensated by the equilibrium flux $J_{B \to A}$  from B to A.  As follows from Eq.~\ref{detailed balance}, we then write\cite{hill_free_1989}
\begin{equation} \label{fluxOverPop}
    k_{A\to B} = \frac{J_{A \to B}}{P_A}.
\end{equation}
 Given our definition of the ``A'' ensemble as all the states with $x<0$, the equilibrium flux from A to B is easily computed.  While the equilibrium values of $P_A$ and $P_B$ are insensitive to the precise choice of the boundary between A and B, the flux depends strongly on it\cite{makarov_single_2015}. Even more disconcertingly, this flux is infinite under the assumption of  diffusive dynamics\footnote{Diffusive dynamics, with a constant diffusivity $D$,  corresponds to the case of the overdamped Langevin equation where the inertial term $m d^2 x/dt^2$ is omitted from Eq.~\ref{Langevin} (see, e.g., ref.\cite{elber_molecular_2020}}. Indeed, in this case the evolution of the probability density $p(x,t)$ of finding the system at point $x$ at time $t$ is governed by the Smoluchowski equation, 
 \begin{equation} \label{smoluchowsky}
\frac{\partial p}{\partial t} = -\frac{\partial J}{\partial x} \equiv \frac{\partial}{\partial x} De^{-\beta U(x)} \frac{\partial}{\partial x} e^{\beta U(x)} p,
\end{equation}
where $D=D(x)$ is the particle's diffusivity and where $J$ is the flux evaluated at point $x$ at time $t$. If we choose  $p(x)\propto \theta(-x)$, where the Heaviside function $\theta(-x)$ selects the particles with $x<0$, then computing the flux $J_{A\to B}$ involves differentiation of this function with respect to $x$ and thus behaves as the delta-function, being singular at $x=0$.   This paradoxical behavior of the flux is a reflection of the famous ``recrossing'' problem: a Brownian particle starting at $x=0$ will typically cross and recross the origin infinite number of times before departing from this point. If each crossing counts as transition between A and B, the resulting flux and thus the estimated transition rate coefficient is infinite!   This is not to say that an experimentalist observing a Brownian particle in a bistable optical trap, or a chemist studying an isomerization reaction, will measure an infinite isomerization rate $k_{A\to B}$. In fact, from an experimental point of view, there is very little controversy about the value of   $k_{A\to B}$, which happens to be finite and quite robust with respect to how it is defined or measured. So the problem must be with how we decided which microscopic states are assigned to the A and B states. 

A physically more meaningful definition of the reactive flux involves placing two boundaries, $a$ and $b$, located to the left and to the right of the barrier top (Fig.~\ref{fig:mapping}, bottom).  We then consider {\it transition paths}\cite{hummer_transition_2004}, trajectory segments that leave the boundary $a$ and proceed directly to the boundary $b$ without returning to $a$. The flux of such trajectories, $J_{a\to b}$, in general depends on the chosen boundaries.  However, when (1) the barrier separating the two minima is much greater than the thermal energy  $k_B T$ and (2) moreover, the boundaries are placed such that $U(0)-U(a) \gg k_B T$, $U(0)-U(b)\gg k_B T$, then this flux becomes essentially independent of the boundaries, and we identify it with the equilibrium reactive flux, 
\begin{equation} \label{reactive flux}
J_{A\to B}\approx J_{a\to b}
\end{equation}
We now introduce a mapping of the equilibrium ensemble onto reactant  (A) and product (B) ensemble members that is consistent with Eq.~\ref{reactive flux} in that it gives a flux of transition paths that is independent of the point $x$, $a<x<b$, where it is measured\cite{vanden-eijnden_exact_2009,warmflash_umbrella_2007}. For a trajectory that passes through a point $x$ within the interval $a<x<b$, we ask whether  it {\it most recently} crossed  the interval boundary  $a$ or $b$. If it was $a(b)$, we identify it as belonging to A(B).  The result is illustrated in Fig. \ref{fig:mapping}, bottom, where red/blue points are classified as those corresponding to A and B. Importantly, with this definition, belonging to A- or B-ensemble is not merely a function of the particle's position but rather a property of the trajectory itself.  
   
We now derive two key consequences of this classification that make it consistent with Eq.~\ref{reactive flux}. 
\subsection*{Distributions $p_A(x)$ and $p_B(x)$ }
First, assuming that the dynamics $x(t)$ is ergodic and obeys time reversal symmetry (i.e., in equilibrium), the probability density of A-points in the interval $(a,b)$ is given by
\begin{equation} \label{density of A points}
    p_A (x)=p_{eq,x}(x) \phi(x\to a|b),
\end{equation}
where $\phi(x\to a|b)$ is the splitting probability\cite{gardiner_handbook_1983} that a trajectory starting at $x$ will hit the boundary $a$ before hitting the boundary $b$.  Similarly, the population of the B-members of the ensemble within the interval $(a,b)$ can be written in terms of the splitting probability to reach the boundary $b$ before reaching $a$.
\begin{equation} \label{density of B points}
    p_B (x)=p_{eq,x}(x) \phi(x\to b|a),
\end{equation}
To prove this result, we write, for a small enough interval $\delta x$ and long enough time interval $\tau$, 
\begin{equation} \label{equil prob} 
p_{eq,x}(x)\delta x \approx \frac{1}{\tau} \int_0^{\tau} dth_{(x,x+\delta x)}[y(t)],
\end{equation}
where $h_{(x,x')}(y)$ is the indicator function equal to 1 if $x<y<x'$ and zero otherwise. Moreover, we have
\begin{equation}
p_{A}(x)\delta x \approx \frac{1}{\tau} \int_0^{\tau} dth_{(x,x+\delta x)}[y(t)] H_a[t,\{y(t)\}],
\end{equation}
where the functional $H_a[t,\{y(t)\}] = 1-H_b[t,\{y(t)\}]$ is equal to 1 if, at time $t$, the boundary $a$ was the last one  crossed by the trajectory $y(t)$. The value of this functional is equal to zero otherwise.  Changing the integration variable from $t$ to $t'=\tau-t$ above, we write
\begin{equation}
p_{A}(x)\delta x \approx \frac{1}{\tau} \int_0^{\tau} dt' h_{(x,x+\delta x)}[\tilde{y}(t')] \tilde{H}_a[t',\{\tilde{y}(t')\}],
\end{equation}
where  
\begin{equation}
\tilde{y}(t)=y(\tau-t)
\end{equation}
defines a time-reversed trajectory and where $\tilde{H}_a[t,\{y(t)\}]$  is the time-reversed indicator functional that is equal to one if $a$  is the first boundary encountered by $y(t)$ {\it after} time $t$ and zero otherwise. Time reversal symmetry implies that replacing $y(t)$ with $\tilde{y}(t)$ will preserve the validity of the above expression, 
\begin{equation}
p_{A}(x)\delta x \approx \frac{1}{\tau} \int_0^{\tau} dt h_{(x,x+\delta x)}[y(t)] \tilde{H}_a[t,\{y(t)\}],
\end{equation}
(where the prime in the dummy variable $t'$ was additionally dropped). Finally, we divide this by Eq.\ref{equil prob} to get 
\begin{equation} \label{phi to a}
p_A(x)/p_{eq,x}(x) = \frac{\int_0^{\tau} dt h_{(x,x+\delta x)}[y(t)] \tilde{H}_a[t,\{y(t)\}]}{\int_0^{\tau} dth_{(x,x+\delta x)}[y(t)]} \approx \phi(x\to a|b),
\end{equation}
as the integral in the numerator of Eq. \ref{phi to a} is the joint probability that a trajectory segment starts within the interval $(x,x+\delta x)$ and arrives at the boundary $a$ before arriving at $b$, and so the ratio of the two integrals (in the limit $\tau \to \infty$ , $\delta\to 0$) is the conditional probability that a trajectory starting at $x$ will arrive at $a$ before $b$. This proves Eq.~\ref{density of A points}. 
\subsection*{Preservation of unidirectional flux in the transition region $(a,b)$.}
Second, consider now the A-members of the ensemble situated just to the left of the boundary $b$. These are the trajectories that started at $a$, stayed continuously within the interval $(a,b)$, and are about to end at $b$. By definition, the flux of those is the reactive flux $J_{a\to b}$. 
\begin{figure}
    \centering
    \includegraphics[width=0.75\linewidth]{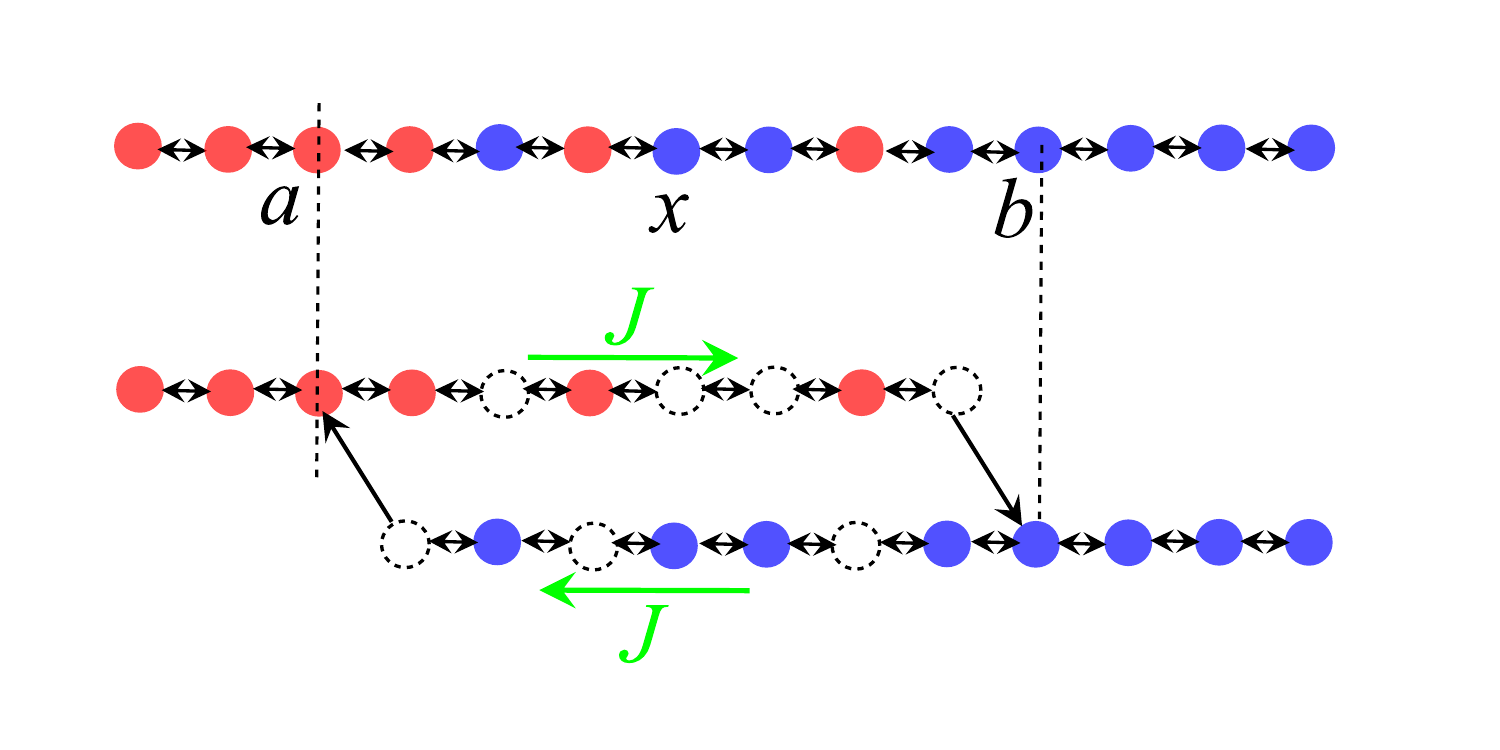}
    \caption{Classifying individual members of the equilibrium ensemble  into two types, A and B. The A-members, indicated in red, most recently came from the boundary $a$.  It is convenient to imagine them occupy an upper track in this diagram reserved just for them. As soon as they reach the boundary $b$, they switch color to blue becoming B-members; they are then transferred to the lower track reserved for them in this picture. As a result, the red A-circles undergo overall drift to the right and the blue B-circles drift to the left. Combining the two tracks results in an equilibrium ensemble with no net flux, but separating the A- and B-points into two tracks results in a nonequilibrium cycle, with a net clockwise flux equal to the unidirectional reactive flux $J_{a\to b}$. Here, for convenience of representation, we assumed discrete hopping dynamics along $x$, which may be thought of as a finite-difference representation of continuous diffusive dynamics.}
    \label{fig:Kramers_cycle}
\end{figure}

If we watch an individual ensemble member, it is in the A-state if the last boundary that it visited was $a$ (Figure~\ref{fig:Kramers_cycle}, red), and it switches to B as soon as it crosses $b$ for the first time (blue). This can be represented by the nonequilibrium cycle shown in  Figure~\ref{fig:Kramers_cycle}, bottom, where the equilibrium ensemble members classified as A (red circles) drift overall to the right along the upper track and those classified as B (blue circles) drift to the left\cite{suarez_simultaneous_2014}. In the steady state, the flux $J\equiv J_{a\to b}$ at any point along the cycle is the same, and thus is independent of $x$ -- the desired property as discussed above. In his paper\cite{kramers_brownian_1940}, Kramers mentions quite briefly and somewhat cryptically that he considers a ``steady-state'' scenario to compute the rate of escape. We speculate that the steady-state scenario depicted in Fig.\ref{fig:Kramers_cycle} is a precise implementation of Kramers' original idea.  We also note that the trick with dividing the equilibrium ensemble into two subensembles undergoing a nonequilibrium cycle of the type shown in Fig.~\ref{fig:Kramers_cycle} turns out to be useful as a computational tool in methods employing enhanced sampling\cite{zuckerman_weighted_2017,suarez_estimating_2016,suarez_simultaneous_2014}. 

\section{Rate coefficients in the case of diffusive dynamics}

\subsection*{Reactive flux and first passage times}
We now apply the above results to calculate the rate of transition from A to B in the case of diffusive dynamics. That is, we assume that the probability density $p(x,t)$ evolves according to Eq.~\ref{smoluchowsky}. The equilibrium reactive flux from $a$ to $b$ is, then
\begin{equation} \label{uni flux diff}
J_{a\to b}=-D(x)e^{-\beta U(x)} \frac{\partial}{\partial x} e^{\beta U(x)} p_A(x) = -D(x) p_{eq,x}(x)\frac{\partial \phi(x\to a|b)}{\partial x}=D(x) p_{eq,x}(x)\frac{\partial \phi(x\to b|a)}{\partial x},
\end{equation}
where Eq.~\ref{density of A points} was used. Although this expression appears to have explicit $x$-dependence, we have already proven above that it is independent of $x$ for $a<x<b$. This can also be seen from an explicit calculation. Indeed, using the known expression for the splitting probability \cite{gardiner_handbook_1983}, 
\begin{equation}\label{splitting}
    \phi(x\to a|b) = \frac{I(x,b)}{I(a,b)}
\end{equation}
with a new function $I(a,x)$ defined by
\begin{equation} \label{integral I}
    I(a,x)= \int_x^b dx e^{\beta U(x)}/D(x),
\end{equation}
we find
\begin{equation}
    \frac{\partial \phi(x\to a|b)}{\partial x}=-\frac{e^{\beta U(x)}}{D(x) I(a,b)},
\end{equation}
which, upon substitution into Eq.\ref{uni flux diff} (and using Eq.~\ref{PMF}), gives
\begin{equation} \label{flux from I}
    J_{a\to b}=\frac{1}{q I(a,b)}.
\end{equation}
The product in the denominator of this equation,
\begin{equation}
   qI(a,b) \equiv  \tau_{RT}, 
\end{equation}
has a physical interpretation as the mean (round trip) time to complete the nonequilibrium cycle of Fig. \ref{fig:Kramers_cycle}. This is further supported by the observation that it can be written as the sum of the mean first passage times from $a$ to $b$ and from $b$ to $a$, for which expressions are known\cite{redner_guide_2001}:
\begin{equation}
    \tau_{RT}=\tau(a\to b)+\tau(b\to a)
\end{equation}
\begin{equation}
    \tau(a\to b)= \int_a^b dy \frac{e^{\beta U(y)}}{D(y)} \int_{-\infty}^y dx e^{-\beta U(x)} 
\end{equation}
\begin{equation}
    \tau(b\to a)= \int_a^b dy \frac{e^{\beta U(y)}}{D(y)} \int_y^{\infty} dx e^{-\beta U(x)} 
\end{equation}

\subsection*{The high barrier limit}
The expressions derived above for the flux $J(a\to b)$ between any two boundaries, as well as for the mean time to complete the round trip from $a$ to $b$ and back to $a$, are exact under the assumption of diffusive dynamics. The simpler, coarse-grained description using the phenomenological first-order rate equations, Eq.~\ref{masterEq}, can only be justified under additional assumptions stated in the discussion of  Eq.~\ref{reactive flux}: the barrier separating the domains A and B must be significantly greater than the thermal energy,  and the boundaries $a$ and $b$ must be suitably chosen, such that the flux $J(a\to b)$ is approximately independent of the boundaries. Indeed, for a sufficiently high barrier and for the boundaries chosen to be far enough from the barrier top, the integral 
$$I(a,b)= \int_a^b dx e^{\beta U(x)}/D(x)$$
is dominated by the integration near the barrier top, where the integrand is close to its maximum value. For example, assuming position-independent diffusivity $D(x)=D$ and quadratic shape of the potential near $x=0$, $U(x)\approx U(0)+U''(0)x^2/2$,  we can estimate this integral as
\begin{equation}
    I(a,b)= \int_a^b dx e^{\beta U(x)}/D(x) \approx D^{-1} \int_{-\infty}^{\infty} dx e^{\beta U(0)} e^{-\beta |U''(0)|x^2/2}=D^{-1} e^{\beta U(0)} \left( \frac{2\pi}{\beta |U''(0)|}\right )^{1/2}, 
\end{equation}
which is independent of $a$ and $b$. Now using Eq.~\ref{fluxOverPop} and Eq.~\ref{flux from I} for the flux, we get
\begin{equation} \label{high barrier rate}
k_{A\to B} = \frac{1}{q P_A I(a, b)},
\end{equation}
where (cf. Eq.~\ref{equilPop})
\begin{equation} \label{pA Integral}
qP_A = \int_{-\infty}^{0} dx e^{-\beta U(x)}
\end{equation}
is dominated by the vicinity of the potential well corresponding to the state A. To estimate it analytically, we write 
\begin{equation}
    U(x)\approx U(x_A)+\frac{1}{2} U''(x_A) (x-x_A)^2,
\end{equation}
substitute this into Eq.~\ref{pA Integral} and extend the upper integration limit to infinity, which gives
\begin{equation}
    qP_A \approx e^{-\beta U(x_A)} \left( \frac{2\pi}{\beta U''(x_A)}\right )^{1/2}.
\end{equation}
Using this with Eq.~\ref{high barrier rate} gives
\begin{equation} \label{Kramers overdamped}
    k_{A\to B} = \frac{D\beta}{2\pi} \sqrt {|U''(0)| U''(x_A)} e^{-\beta [U(0)-U(x_A)]},
\end{equation}
which is Kramers' estimate for the transition rate from A to B in the overdamped limit (i.e., for diffusive dynamics).

\section{Rate coefficients in the case of Langevin dynamics with intermediate friction.}
\subsection*{Reactive flux}
We now consider the case of the dynamics governed by the Langevin equation, Eq.~\ref{Langevin}.
The results of the previous Section correspond to the overdamped limit, where the inertial term in Eq.~\ref{Langevin} is negligible. Here, this limit is no longer assumed. To write the expression for the reactive flux, $J_{A\to B}$, it is now necessary to consider the phase space $(x,v)$  of the system, where $v$ is the velocity.  The equilibrium phase-space density is given by the Boltzmann distribution, 
\begin{equation} \label{Boltzman phase space}
    p_{eq}(x,v)=p_{eq,x}(x) p_{eq,v}(v)= \frac{1}{q} e^{-\beta U(x)} p_{eq,v}(v),
\end{equation}
where 
\begin{equation} \label{Boltzmann velocity}
p_{eq,v}(v)=\sqrt{\frac{\beta m}{2\pi}}e^{-\beta m v^2/2}
\end{equation}
is Maxwell's velocity distribution and $q$ is given by Eq.~\ref{partition function}. We now extend the arguments from the preceding section to write the distributions for the A -- subensemble of the equilibrium ensemble as 
\begin{equation} \label{densisty of A points in phase space}
p_A (x,v)=p_{eq}(x,v) \phi((x,-v)\to a|b),
\end{equation}
where $\phi((x,v)\to a|b)$ denotes the splitting probability to get to the boundary $a$ before crossing the boundary $b$, having started at point $x$ with velocity $v$\footnote{As further discussed in Appendix A, this expression for $p_A(x)$ is equivalent to the unnamed equation preceding Eq. 18  of the Kramers paper\cite{kramers_brownian_1940}, where $p_{eq}$ is the same as $\rho$  and where the splitting probability $\phi$ is the function $\zeta$, thus clarifying precisely how the steady-state scenario envisioned by Kramers can be constructed.}. The derivation of this equation is analogous to that of Eq.~\ref{density of A points} and uses the same time-reversal symmetry argument, which also necessitates that the velocity $v$ is reversed in the splitting probability appearing in Eq.~\ref{densisty of A points in phase space}.    

We now write the unidirectional flux from $a$ to $b$ as
\begin{equation} \label{a to b flux phase space 0}
    J_{a\to b} = \int_{-\infty}^{\infty} p_A(x,v) v dv=\int_{-\infty}^{\infty} p_{eq}(x,v) \phi((x,-v)\to a|b) v dv. 
\end{equation}
Using the symmetry $p_{eq}(x,v)=p_{eq}(x,-v)$ and the fact that $\phi((x,v)\to a|b)= 1-\phi((x,v)\to b|a)$, where $\phi((x,v)\to b|a)$ is the splitting probability of getting to $b$ before crossing $a$ having started at a phase-space point $(x,v)$, we can rewrite the flux, in an intuitively appealing way,  as
\begin{equation} \label{a to b flux phase space}
    J_{a\to b} = \int_{-\infty}^{\infty} p_{eq}(x,-v) [1-\phi((x,-v)\to b|a)] v dv =\int_{-\infty}^{\infty} p_{eq}(x,v) \phi((x,v)\to b|a) v dv, 
\end{equation}
where the identity of the total equilibrium flux to zero, $\int_{-\infty}^{\infty} dv p_{eq}(x,v) v=0$, was also used.  Using Eqs.~\ref{Boltzman phase space} and \ref{Boltzmann velocity} and integrating by parts, this may be rewritten as
\begin{equation} \label{a to b flux phase mod}
    J_{a\to b} = \frac{p_{eq,x}(x)}{\beta m}\int_{-\infty}^{\infty} p_{eq,v}(v) \frac{\partial \phi((x,v)\to a|b)}{\partial v}  dv, 
\end{equation}
\subsection*{Parabolic barrier approximation and Kramers formula for high barriers}
Although a general analytical solution for the splitting probability  $\phi((x,v)\to a|b)$ is unknown, an approximate solution is known for a bistable system of Fig.~\ref{fig:mapping} when the barrier is sufficiently high, and when the parabolic approximation, 
\begin{equation} \label{parabolic barrier}
    U(x)\approx U(0)+\frac{1}{2}U''(0)x^2,
\end{equation}
($U''(0)<0$) can be used for the barrier shape in the vicinity of its top.  In this case, one finds (see Appendix A)
\begin{equation}\label{splitting prob phase space}
    \phi((x,v)\to a|b) \approx \frac{1}{\sqrt{\pi}} \int_{-\infty}^{\sqrt{\omega \kappa/(2 D_v)}(v+\omega x/\kappa)} ds e^{-s^2/2},
\end{equation}
where
\begin{equation}
    \omega = \sqrt{-U''(0)/m}
\end{equation}
is the upside down barrier frequency, 
\begin{equation}
    D_v=\frac{\gamma}{\beta m^2}, 
\end{equation}
is the diffusivity along the velocity coordinate, and 
\begin{equation} \label{Kramers kappa}
    \kappa = \sqrt{1+\frac{\gamma^2}{4m^2\omega^2}}-\frac{\gamma}{2 m \omega}
\end{equation}
is the Kramers transmission coefficient \cite{kramers_brownian_1940}. Then Eq.~\ref{a to b flux phase mod} gives
\begin{equation} \label{a to b flux moderate friction}
    J_{a\to b} = \frac{\kappa}{q\sqrt{2\pi\beta m}}e^{-\beta U(0)}. 
\end{equation}
Substituting this into Eq.\ref{fluxOverPop}, we obtain Kramers' famous result,
\begin{equation} \label{Kramers moderate fric}
    k_{A\to B} = \kappa k_{A\to B}^{TST}
\end{equation}
where 
\begin{equation} \label{TST}
     k_{A\to B}^{TST}=\frac{1}{q P_A \sqrt{2\pi\beta m}}e^{-\beta U(0)}
\end{equation}
is independent of the friction coefficient and known as the transition-state-theory (TST) estimate for the rate coefficient. Transition state theory\cite{peters_reaction_2017} is an approximation to the Kramers result obtained by setting $\kappa=1$. This is equivalent to setting $x=0$ in Eq.\ref{a to b flux phase space} and replacing the splitting probability with the Heaviside step function, 
\begin{equation}
    \phi((0,v)\to b|a) = \theta(v).  
\end{equation}
Physically, this amounts to estimating the flux at the barrier top  assuming that every trajectory crossing this point from left to right (i.e., with $v>0$) will end up hitting the boundary $b$ before hitting $a$ -- barrier recrossings are ignored!

\section{Reaction rates and correlation functions}
\subsection*{Transition rates and conditional probabilities: phenomenological two-state dynamics}
Within the phenomenological model of Eq.~\ref{AB} the system's dynamics is governed by the master equation, Eq.~\ref{masterEq}, which can be written in the following form:
\begin{equation} \label{masterEq 1}
\frac{d \mathbf{P}}{dt} = \mathbf{K} \mathbf{P},
\end{equation}
where $\mathbf{P}=\begin{pmatrix}
P_A(t) \\
P_B(t) \\
\end{pmatrix}$ is the column vector whose components are the probabilities to find the system in states A and B at time $t$, and where 
\begin{equation} \label{rate Matrix}
   \mathbf{K}= (\begin{smallmatrix}
  -k_{A\to B} & k_{B\to A}\\
  k_{A\to B} & -k_{B\to A}
\end{smallmatrix})
\end{equation}
is the rate (generator) matrix for the two-state system, Eq.~\ref{AB}.  The solution of Eq.~\ref{masterEq 1} is 
$$
\mathbf{P}(t) = e^{\mathbf{K}t} \mathbf{P}(0),
$$
from which it follows that the matrix elements of $e^{\mathbf{K}t}$ can be interpreted as conditional probabilities (Green's functions): For example, $P(B,t|A,0)=(e^{\mathbf{K}t})_{21}$ and $P(A,t|A,0)=1-P(B,t|A,0)=(e^{\mathbf{K}t})_{11}$ are the conditional probabilities of observing the system, respectively, in states $B$ and $A$ at time $t$ given that it was in $A$ at time zero. The reaction rate coefficient $k_{A\to B}$ is related to  such probabilities\cite{makarov_single_2015}:
\begin{equation} \label{rate from cond prob}
    k_{A\to B}=\lim_{t\to 0}\frac{dP(B,t|A,0)}{dt}=-\lim_{t\to 0}\frac{dP(A,t|A,0)}{dt}.  
\end{equation}
This explains the meaning of $k_{A\to B}$ as the transition probability from A to B  ``per unit time''. In this picture, the equilibrium unidirectional flux $J_{A\to B}$ is (cf. Eq.~\ref{detailed balance}))
\begin{align} \label{from joint probs to fluxes}
\begin{split}
J_{A\to B}=J_{B\to A}&=\lim_{t\to 0} \frac{dP(B,t;A,0)}{dt}=-\lim_{t\to 0} \frac{dP(A,t;A,0)}{dt}\\
&=\lim_{t\to 0} \frac{dP(A,t;B,0)}{dt}=-\lim_{t\to 0} \frac{dP(B,t;B,0)}{dt}.
\end{split}
\end{align}
Here $P(B,t;A,0)= P(B,t|A,0) P_A$ and $P(A,t;A,0)= P(A,t|A,0) P_A$ are the joint probabilities  of observing the system in $A$ at $t=0$ and in $B$ or $A$ at time $t$, and $P_A$ is the equilibrium probability to be in $A$. The joint probabilities $P(A,t;B,0)= P(A,t|B,0) P_B$ and $P(B,t;B,0)= P(B,t|B,0) P_B$ are defined similarly. To be specific, let us focus on $P(B,t;B,0)$, which can also be written as 
\begin{equation}
    P(B,t;B,0)= (e^{\mathbf{K}t})_{22} P_B, 
  \end{equation}
Using this we can also write
\begin{equation} \label{PBB TSS}
    P(B,t;B,0)=P_B(P_B+P_A e^{\epsilon_1 t}),
\end{equation}
where 
\begin{equation} \label{epsilon1 from rates}
  \epsilon_1=-(k_{A\to B}+k_{B\to A})
\end{equation}
is the non-zero eigenvalue of the rate matrix, Eq.~\ref{rate Matrix} (equal to the relaxation rate of the two-state system taken with the minus sign).  The joint probability $P(B,t;B,0)$, therefore, can be used to determine the reaction rate coefficients $k_{A\to B}$ and $k_{B\to A}$.

Consider now the eigenvectors of $\mathbf{K}$. The vector $$\mathbf{v}_0 = \begin{pmatrix}
P_A \\
P_B \\
\end{pmatrix},$$ with its components  equal to the equilibrium probabilities  to find the system in states A and B corresponds to zero eigenvalue, consistent with the notion that if one starts with the populations described by $\mathbf{P}(0)=\mathbf{v}_0$, then the probabilities to find the system in each state, which evolve according to Eq.~\ref{masterEq 1}, will stay constant, $\mathbf{P}(t)=\mathbf{P}(0)$. One can easily check that the eigenvector $\mathbf{v}_1=\begin{pmatrix}
v_A \\
v_B \\
\end{pmatrix}$ corresponding to the nonzero eigenvalue $\epsilon_1$ satisfies the relationship
\begin{equation} \label{TSS_orthogonality}
    v_A +v_B =0,
\end{equation}
and thus is orthogonal to $\mathbf{v}_0$. We will use this type of orthogonality condition below to determine approximate time evolution of more complicated dynamics governed by the Smoluchowski or Klein-Kramers equations.

\subsection*{Diffusive dynamics}
For diffusive dynamics along a continuous reaction coordinate $x$, the above joint probability can be expressed as an autocorrelation function of the indicator function $h_{B}(x)\equiv \theta( x)$, which is equal to 1 when $x$ is to the right of the barrier top and zero otherwise: 
\begin{equation} \label{correlator from GF}
    P(B,t;B,0)=\langle h_B[x(t)] h_B[x(0)]\rangle = \int_{0}^{\infty}dx \int_{0}^{\infty}dx_0 G(x,t|x_0,0) p_{eq,x}(x_0),
\end{equation}
where $G(x,t|x_0,0)$ is the propagator (Green's function) associated with the Smoluchowski equation (i.e., the solution of Eq.\ref{smoluchowsky} with the initial condition $p(x,0)=\delta(x-x_0)$).  Writing this equation in the form
\begin{equation}\label{sm Eq}
    \frac{\partial G}{\partial t}=\hat{L}G, 
\end{equation}
where the Smoluchowski time-evolution operator is given by
\begin{equation}\label{smoluchowski operator}
    \hat{L}=\frac{\partial}{\partial x}D(x) e^{-\beta U(x)} \frac{\partial}{\partial x} e^{\beta U(x)}, 
\end{equation}
reveals the analogy between the Smoluchowski time evolution and that of Eq.~\ref{masterEq 1}. The propagator can  be written as a spectral expansion, 
\begin{equation} \label{GF spectral}
    G(x,t|x_0,0)=p_{eq,x}(x)+\sum_{n=1}^{\infty}\psi_n(x)\psi_n(x_0)\frac{1}{p_{eq,x}(x_0)}e^{\epsilon_n t}
\end{equation}
where $\psi_n$ and $\epsilon_n$ are the eigenfunctions and the eigenvalues of the Smoluchowski time-evolution operator, which are the solutions of the equation
\begin{equation}\label{eig problem SM}
    \hat{L}\psi_n(x)=\epsilon_n\psi_n(x)
\end{equation}
Here, the largest eigenvalue is  $\epsilon_0=0$, with the eigenfunction corresponding to the equilibrium distribution,  $\psi_0(x)=p_{eq,x}(x)$. All other eigenvalues describe relaxation of the system: accordingly, they are negative and can be arranged in descending order  $0>\epsilon_1 >\epsilon_2 >\epsilon_3\ldots.$  

Substituting Eq.~\ref{GF spectral} into Eq.~\ref{correlator from GF}, we obtain a spectral expansion of the joint probability,
\begin{equation} \label{PBB spectral}
    P(B,t; B,0)=P_{B}^2+\sum_{n=1}^{\infty}\left [\int_0^\infty dx\psi_n(x)\right]^2e^{\epsilon_n t}.
\end{equation}
At $t\gg |\epsilon_2|^{-1}$ this can be approximated by\footnote{Replacing the literal $t\to 0$ limit in Eq.~\ref{from joint probs to fluxes} with ``intermediate'' times $|\epsilon_2|^{-1} \ll t \ll |\epsilon_1|^{-1}$ is the essential step in the computation of the reaction rate using the correlation function formalism\cite{chandler_notitle_1978,bennett_molecular_1977, hanggi_50_1990}. This is possible when the barrier is high enough that the intra-well relaxation  (governed by $\epsilon_i, i\ge 2$) is much faster than inter-well relaxation. In this context,   $|\epsilon_2|^{-1} $ is usually referred to as the molecular time.} 
\begin{equation} \label{PBB from greens function}
    P(B,t; B,0)\approx P_{B}^2+\left [\int_0^\infty dx\psi_1(x)\right]^2e^{\epsilon_1 t},
\end{equation}
which appears similar to Eq.~\ref{PBB TSS} written for a two-state system.  

To explore this analogy further, we focus on the slowest relaxation mode, with the eigenfunction $\psi_1(x)$ and the eigenvalue  $\epsilon_1$. In the presence of a sufficiently high barrier separating two potential wells, this mode describes equilibration between the two wells via activated barrier crossing. The corresponding relaxation time is much longer than the relaxation time within the potential wells, and thus $|\epsilon_1|$ is much smaller than $|\epsilon_2|, |\epsilon_3| \ldots$.  Ref.~\cite{berezhkovskii_diffusive_2021}, showed that, in this regime, the slowest relaxation mode can be estimated using the following line of reasoning. First, recall that, in the barrier region $a\le x\le b$ , the equilibrium distribution can be decomposed into the reactant and product components as
\begin{equation} \label{ground state decomposition}
    p_{eq,x}(x)=\psi_0(x)= p_A(x)+p_B(x)\equiv p_{eq,x}(x)\phi(x\to a|b)+p_{eq,x}(x) \phi(x\to b|a)
\end{equation}
Outside the barrier region, we require that $\psi_0(x)$ simply coincides with $p_{eq,x}(x)$. That is, $p_A(x)=p_{x,eq}(x)$,  $p_B(x)=0$  for $x<a$  and $p_B(x)=p_{x,eq}(x)$,  $p_A(x)=0$ for $x>b$. With this convention, Eq.~\ref{ground state decomposition} decomposes the ``ground state'' \footnote{While, here, this is the state with the largest eigenvalue, we use the terms ``ground'' and ``first excited'' states to emphasize the analogy with quantum mechanics. It will be the ground state of the operator -$\hat{L}$} eigenfunction of the Smoluchowski operator into two functions localized in the potential wells.
The ``first excited state'' must be orthogonal to $\psi_0$,  
$$
(\psi_0,\psi_1)=\int_{-\infty}^{\infty}dx \psi_0(x)\psi_1(x)/p_{eq,x}(x)=\int_{-\infty}^{\infty}dx \psi_1(x)=0,
$$
and it should be normalized
$$
(\psi_1,\psi_1)=\int_{-\infty}^{\infty}dx \psi_0^2(x)/p_{eq,x}(x) = 1.
$$
Here 
$$(f,g)=\int_{-\infty}^{\infty}dx f(x) g(x)/p_{eq,x}(x) $$
denotes the scalar product of the functions $f(x)$ and $g(x)$.  In the spirit of constructing molecular orbitals as linear combinations of atomic ones, one can propose approximating $\psi_1$ as a linear combination of the form
\begin{equation} \label{ansatz}
\psi_1(x)\approx \tilde{\psi}_1(x)=c_{A}p_A(x)-c_{B} p_B (x),
\end{equation}
where the coefficients $c_{A}$ and $c_{B}$ are determined by the above orthonormality conditions. In particular, orthogonality requires that 
$$
c_{A}P_A - c_{B}P_B=0,
$$
a condition analogous to the orthogonality condition stated, for a two-state system, in Eq.~\ref{TSS_orthogonality}. Combined with the normalization condition, this gives
\begin{equation} \label{lcao coefficients}
    c_{A}=\sqrt{\frac{P_B}{P_A}}, \quad c_{B}=\sqrt{\frac{P_A}{P_B}}.
\end{equation}
\begin{figure}
    \centering
    \includegraphics[width=0.75\linewidth]{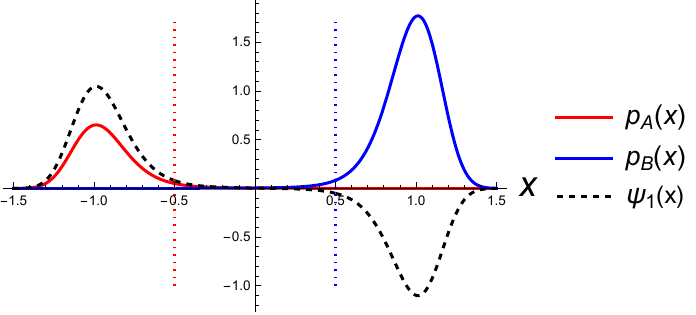}
    \caption{Approximating the first relaxation mode (first eigenfunction) of the Smoluchowski equation as a linear combination of functions $p_A(x)$ and $p_B(x)$ localized in potential wells, Eq.~\ref{ansatz}. These functions were computed numerically for $U(x)=5 k_B T \left[  (x^2-1)^2 -0.5x\right ]$. We emphasize that the approximate $\tilde{\psi}_1(x)$ is visually indistinguishable from the exact eigenfunction $\psi_1(x)$ computed numerically. The locations of the boundaries $a$ and $b$ used in this example are shown as vertical dotted lines. }
    \label{fig:relaxation mode}
\end{figure}
The quality of the approximation of Eq.~\ref{ansatz} is illustrated in Fig.~\ref{fig:relaxation mode}, and some of the technical subleties of using this function to approximate the true eigenfunction are further highlighted in Appendix B. 

Using Eqs. \ref{ansatz} and \ref{lcao coefficients}, it follows immediately that $\left [\int_0^\infty dx\psi_1(x)\right]^2=P_A P_B$, making the expression for the joint probability $P(B,t;B,0)$, Eq.~\ref{PBB from greens function}, identical to that in Eq.~\ref{PBB TSS}\footnote{This shows that the two-state approximation, Eq.~\ref{PBB TSS}, is valid not only because the terms corresponding to the higher-order eigenfunctions decay fast  in Eq.~\ref{PBB spectral}, but also because their {\it amplitude} must be small, since at $t=0$ our approximation already correctly recovers $P(B,0;B,0)\equiv P_B$. This, in turn, explains why nonexponential time dependence of  experimental observables would be difficult to detect even when the time resolution is sufficient for detecting the fast-decaying components of the experimental signal}.  This strongly suggests that the eigenvalue $\epsilon_1$ can be expressed as the sum of the forward and backward transition rates for diffusive barrier crossing, as stated in Eq.~\ref{epsilon1 from rates}. 

Let us show, explicitly, that this is indeed the case.  Given the eigenfunction $\psi_1$, the eigenvalue $\epsilon_1$ is given by
\begin{equation} \label{expected value eps1}
\epsilon_1 = (\psi_1,\hat{L}\psi_1)=\int_{-\infty}^{\infty}dx \psi_1(x) \hat{L}\psi_1(x)/p_{eq,x}(x)
\end{equation}
To evaluate this integral, it is expedient to introduce new functions 
\begin{equation}
    \chi_n(x)=\psi_n(x)/p_{eq}(x),
\end{equation}
which are the eigenfunctions of the adjoint Smoluchowski operator 
\begin{equation}\label{adjoint smoluchowski operator}
    \hat{L}^{\dagger}= [p_{eq,x}(x)]^{-1}\hat{L} p_{eq,x}(x) = e^{\beta U(x)} \frac{\partial}{\partial x} D(x) e^{-\beta U(x)}\frac{\partial}{\partial x} 
\end{equation}
with the same eigenvalues ($\epsilon_n$). In terms of this operator and the new functions, we can also rewrite Eq.~\ref{expected value eps1} as 
\begin{equation} \label{expected value eps1 adjoint}
\epsilon_1 =\int_{-\infty}^{\infty}dx p_{eq,x}(x)\chi_1(x) \hat{L}^{\dagger}\chi_1(x)
\end{equation}
Substituting the definition of the adjoint operator, Eq.~\ref{adjoint smoluchowski operator}, into Eq.~\ref{expected value eps1 adjoint} and integrating by parts, we find
\begin{equation} \label{eps1}
\epsilon_1 =\int_{-\infty}^{\infty}dx D(x) p_{eq,x}(x)\left [\frac{d \chi_1(x)}{dx}\right ]^2 .
\end{equation}
If, instead of the true eigenfunction $\chi_1$, one uses 
$$
\tilde{\chi}_1(x) = \tilde{\psi}_1(x)/p_{eq,x}(x),
$$
then the following estimate is obtained:
\begin{equation} \label{eps1 approx}
\epsilon_1 \approx -\int_{a}^{b}dx D(x) p_{eq,x}(x)\left [\frac{d \tilde{\chi_1}(x)}{dx}\right ]^2 ,
\end{equation}
where we have recognized that the integrand vanishes outside the interval $[a,b]$. Note that, according to Eq.~\ref{ansatz}, we have
\begin{equation} \label{ground state decomposition chi}
   \tilde{\chi}_1(x)=c_A \phi(x\to a|b)-c_B \phi(x\to b|a)
\end{equation}
for $a\le x\le b$. Using Eqs.~\ref{splitting},
 \ref{integral I}, and \ref{lcao coefficients}, we obtain 
 $$
 \frac{d\tilde{\chi}_1}{d x}=\frac{e^{\beta U(x)} }{D(x) I(a,b) \sqrt{P_A P_B}}.
 $$
Substituting this into Eq.~\ref{eps1 approx}, one eventually gets 
\begin{equation} \label{eps1 final}
    \epsilon_1 \approx -\frac{J_{a\to b}}{P_A P_B} = -(\frac{J_{a\to b}}{P_A}+\frac{J_{a\to b}}{P_B})=-(k_{A\to B}+k_{B\to A}),
\end{equation}
where $J_{a\to b}$ is given by Eq.~\ref{flux from I} and where $k_{A\to B}$ and $k_{B\to A}$ are the transition rates in the overdamped case.  This result is identical to the expression in  Eq.~\ref{epsilon1 from rates} obtained using two-state kinetics\cite{berezhkovskii_diffusive_2021}.  It is worth emphasizing that our result is unexpected: If one were to use Eq.~\ref{eps1 final} with the variational estimate, Eq.~\ref{eps1 approx}, to infer the unidirectional flux $J_{a\to b}$ one would recover the result  of Eq.~\ref{flux from I}. But Eq.~\ref{flux from I} is the {\it exact} expression for the equilibrium unidirectional flux between the two boundaries $a$ and $b$. In contrast, the variational ansatz, Eq.~\ref{ansatz}, {\it is not} an exact solution for the eigenfunction $\psi_1$ -- in fact, it fails to solve the eigenvalue equation quite spectacularly, as shown in Appendix B. We do not know whether the identity of the variational estimate of $J_{a\to b}$ and the true flux is a fortuitous coincidence or has a deeper origin. 

To summarize this section, we showed that diffusive dynamics in a double-well potential with a sufficiently high barrier can be mapped onto that of a two-state system, with the inter-state transition rates $k_{A\to B}$ and $k_{B\to A}$  identical to those obtained from the equilibrium reactive flux considerations in Section II. This mapping can be understood from two different perspectives: One emerges when considering the slowest relaxation mode of the Smoluchowski operator and comparing it  with the relaxation rate (equal to the sum of the forward and backward rates) of the two-state system. The other perspective focuses on the short-time behavior of the joint probabilities  of observing the system, e.g., to the left and to the right of the barrier\cite{makarov_single_2015,elber_molecular_2020}  (which can be equivalently thought of as position correlation functions as in some of the formulations of ``exact'' rate theory\cite{bennett_molecular_1977,chandler_notitle_1978,hanggi_50_1990}), that can be obtained using the spectral expansion of the Smoluchowski operator.  Remarkably, both of  these perspectives converge on the answer given by the equilibrium reactive flux considerations! 

\subsection*{Langevin dynamics}

The analysis of the previous section can be generalized for the case of dynamics in phase space. Specifically, Eq.~\ref{correlator from GF} now takes the form 
\begin{equation} \label{correlator from KK}
    P(B,t;B,0)=\langle h_B[x(t)] h_B[x(0)]\rangle = \int_{0}^{\infty}dx \int_{0}^{\infty} dx_0 \int_{-\infty}^{\infty} dv \int_{-\infty}^{\infty} dv_0 G(x,v,t|x_0,v_0,0) p_{eq}(x_0,v_0),
\end{equation}
where the system's propagator in phase space, $G(x,v,t|x_0,v_0,0)$, obeys the Klein-Kramers  (Fokker-Planck) equation\cite{kramers_brownian_1940,risken_fokker-planck_1996}
\begin{equation}
    \frac{\partial G}{\partial t}=\hat{L}_{KK}G,
\end{equation}
with the Klein-Kramers operator $\hat{L}_{KK}$ given by
\begin{equation} \label{LKK}
    \hat{L}_{KK}=-v\frac{\partial}{\partial x}+\frac{U'(x)}{m}\frac{\partial}{\partial v}+D_v \frac{\partial}{\partial v} e^{-\beta mv^2/2}\frac{\partial}{\partial v} e^{\beta mv^2/2}.
\end{equation}
Again, we write the propagator as a spectral expansion, 
\begin{equation} \label{KK spectral}
    G(x,v,t|x_0,v_0,0)=p_{eq}(x,v)+\sum_{n=1}^{\infty}\psi_n(x,v)\psi_n(x_0,-v_0)\frac{1}{p_{eq}(x_0,v_0)}e^{\epsilon_n t},
\end{equation}
where $\psi_n$ and $\epsilon_n$ solve the eigenvalue equation
\begin{equation}\label{eig problem KK}
    \hat{L}_{KK}\psi_n(x,v)=\epsilon_n\psi_n(x,v).
\end{equation}
Once again, the largest in magnitude eigenvalue is  $\epsilon_0=0$, with the eigenfunction equal  to the equilibrium distribution,  $\psi_0(x,v)=p_{eq}(x,v)$. The remaining eigenvalues have negative real parts and  can be arranged in descending order  $0>\mathrm{Re} (\epsilon_1 )>\mathrm{Re} (\epsilon_2) >\mathrm{Re} (\epsilon_3)\ldots$ . 

Substituting Eq.\ref{KK spectral} into Eq.\ref{correlator from KK}, we find
\begin{equation}
    P(B,t; B,0)=P_{B}^2+\sum_{n=1}^{\infty}\left [\int_0^\infty dx\int_{-\infty}^{\infty} dv\psi_n(x,v)\right]^2e^{\epsilon_n t},
\end{equation}
which can be approximated, at $t\gg |\epsilon_2|^{-1}$,  by 
\begin{equation} \label{PBB from KK}
    P(B,t; B,0)\approx P_{B}^2+\left [\int_0^\infty dx\int_{-\infty}^{\infty} dv \psi_1(x,v)\right]^2e^{\epsilon_1 t},
\end{equation}
 which analogous to Eq.~\ref{PBB from greens function}
found in the case of diffusive dynamics.  

By analogy with the case of diffusive dynamics, the eigenfunction $\psi_1(x,v)$corresponding to the slowest relaxation mode can be approximated as a linear combination of the reactant and product distributions using an ansatz of the form
\begin{equation} \label{ansatz phase space}
\psi_1(x,v)\approx \tilde{\psi}_1(x,v)=c_{A}p_{eq}(x,v) \phi((x,-v)\to a|b)-c_{B} p_{eq}(x,v) \phi((x,-v)\to b|a),
\end{equation}
and the conditions of orthogonality of $\psi_1(x,v)$ to $\psi_0(x,v)=p_{eq}(x,v)$, $(\psi_1,\psi_0)=0$, and normalization, $(\psi_1,\psi_1)=1$. Here, the scalar product of two functions, $f(x,v)$ and $g(x,v)$, is defined as 
$$
(f,g) = \int_{-\infty}^{\infty}dx \int_{-\infty}^{\infty}dv f(x,-v) g(x,v)/p_{eq}(x,v).
$$
This  leads to the same values for the coefficients $c_{A,B}$ as in the diffusive case, Eq.~\ref{lcao coefficients}.  Then one finds that the expression in the square brackets in Eq.~\ref{PBB from KK} is equal to $-\sqrt{P_A P_B}$, and this equation is reduced to Eq.~\ref{PBB TSS}. Moreover, following essentially the same steps as in the diffusive case but using the Klein-Kramers operator instead of the Smoluchowski operator, with its corresponding eigenfunctions (Appendix C), one obtains a variational estimate for $\epsilon_1$ given by Eq.~\ref{eps 1 variational Langevin}. If we now interpret this estimate, as before, as the sum of the forward and backward coefficients
\begin{equation} \label{Langevin eps1}
\epsilon_1 \approx  (\tilde{\psi}_1,\hat{L}_{KK}\tilde{\psi}_1)=-(k_{A\to B}+k_{B\to A})=-(\frac{J_{a\to b}}{P_A}+\frac{J_{a\to b}}{P_B}),
\end{equation}
 then we get an estimate for the unidirectional flux $J_{a\to b}$ given by Eq.~\ref{flux from eps1 Langevin}. Although it appears different from the flux expression obtained by direct calculation, Eq. \ref{a to b flux phase mod}, one can show that it leads to the same expression for the reactive flux, Eq.~\ref{a to b flux moderate friction}, and thus the same Kramers formula for the transition rate, Eq.~\ref{Kramers moderate fric}, under the usual assumption of a parabolic barrier of a height substantially exceeding the thermal energy.   

\section{Conclusions}
The mapping of continuous dynamics onto rate equations of chemical kinetics (known as the master equation in physics)  is encountered in many problems in physics and chemistry.  Starting from the Arrhenius law discovered in the 19-th century, chemists have long recognized that the rate coefficients appearing in these equations show a characteristic Arrhenius-like temperature dependence, Eq.~\ref{Arrhenius}; their dependence on the microscopic details of the molecular dynamics is encoded by just two numbers, the activation barrier and  the preexponential factor in the Arrhenius law. Kramers' derivation of the Arrhenius law, based on the simple model of escape of a Brownian particle from a potential well, provided an explanation why solution-phase reactions are usually much slower than their gas-phase counterparts and predicted inverse proportionality of the prefactor to the solvent viscosity in the overdamped case. These insights were crucial to our understanding of  biochemical phenomena occurring in living systems,  as all of them take place in water. As a result, Kramers' model was widely adopted to describe phenomena such as biomolecular folding\cite{klimov_viscosity_1997,socci_diffusive_1996}, particularly as a means for interpreting experimental studies in terms of a small number of experimentally determinable parameters.   Two parallel developments that took place over the last two decades  warrant revisiting the foundations of Kramers' theory today. One is the development of computational methods such as Markov-state models\cite{pande_everything_2010} and milestoning\cite{elber_perspective_2016}, which map continuous molecular dynamics onto discrete-state processes with the goal of both increasing computational efficiency and  elucidating the mechanisms of complex biophysical phenomena\cite{piana_atomic-level_2013}.  The other one is the progress in single-molecule techniques, which allowed probing the biochemical dynamics beyond the chemical kinetics description and characterizing barrier crossing dynamics experimentally\cite{chung_fast_2014,chung_protein_2018,sturzenegger_transition_2018,hoffer_probing_2019,zijlstra_transition_2020}. Those, in particular, allowed direct tests of the Kramers model in application to processes such as biomolecular folding\cite{woodside_determining_2014,neupane_protein_2016,neupane_testing_2018,neupane_measuring_2018,hoffer_measuring_2019} and stimulated developments in chemical dynamics that go beyond determining the transition rates\cite{autieri_dominant_2009,zhang_transition-event_2007,chaudhury_harmonic_2010,berezhkovskii_single-molecule_2018,berezhkovskii_communication_2018,makarov_shapes_2015,carlon_effect_2018,laleman_transition_2017,kim_mean_2015,daldrop_transition_2016,koehl_realistic_2026}.   

Here we have shown that dividing the equilibrium ensemble of a bistable system into two nonequilibrium sub-ensembles corresponding to the reactant and products results in a mapping of the continuous dynamics onto two-state kinetics that  corresponds to phenomenological chemical kinetics. This mapping is closely related to that provided by a collection of  methods that purport to identify reactive events while eliminating spurious transitions or recrossings \cite{vanden-eijnden_exact_2009,vanden-eijnden_transition_2006, warmflash_umbrella_2007, hummer_transition_2004, best_reaction_2005, elber_perspective_2016,berezhkovskii_committors_2019}.   Importantly, the equations describing the equilibrium unidirectional fluxes between the boundaries defining the ``reactant'' and ``product'' states (Eqs.~\ref{uni flux diff} and \ref{flux from I} in the diffusive case and Eq.~\ref{a to b flux phase mod} in the Langevin dynamics case) are exact for any potential $U(x)$ even when the system's dynamics cannot be well approximated by first order chemical kinetics. More precisely, these equations give the {\it exact} flux of transition paths (i.e. trajectory segments traversing the interval $(a,b)$ from $a$ to $b$ without leaving it) for two arbitrary boundaries $a$ and $b$, but only when the potential $U(x)$  has a sufficiently high barrier between the two boundaries can this flux be interpreted as a reactive flux between two chemical species A and B.  

In addition to this exact approach to calculating the fluxes, we have also explored two closely connected approximations, one based on a spectral representation of correlation functions\cite{bennett_molecular_1977,chandler_notitle_1978}/joint probabilities\cite{elber_molecular_2020} and the other based on a variational estimate of  the slowest relaxation rate (identified with the sum of the forward and backward rate coefficients). These methods are approximate and assume a spectral gap allowing one to neglect the contributions from faster relaxation rates into the system's dynamics at times that are not too short. Moreover,  the slowest relaxation rate cannot be evaluated analytically, and a variational ansatz is used instead\cite{berezhkovskii_diffusive_2021}. Surprisingly, these approximate methods nevertheless produce expressions for the reactive fluxes that are essentially the same (identical in the diffusive case) as the exact ones.   

In principle, the splitting-probability/transition-path-based mapping onto a discrete system explored here is not unique: simply dividing the configuration or phase space into two domains, A and B, results in a different mapping, where the kinetics of transitions between A and B is not Markovian\cite{berezhkovskii_barrier_2025}.  This non-Markov process has interesting properties; for example, the average time spent in states A and B is not $1/k_{A\to B}$  but rather $1/k_{A\to B}^{TST}$, thereby giving the transition-state theory a precise experimental meaning\cite{berezhkovskii_barrier_2025,anslyn_transition_2026}.  The transition-path-based mapping, however, most closely corresponds to the chemical view of the reaction and -- from a computational perspective -- also has many advantages (see, e.g., refs.\cite{d_hartich_emergent_2021,blom_milestoning_2024} for applications of this idea to stochastic thermodynamics).

Looking forward, while Kramers' model provided a satisfying qualitative picture of complex biomolecular processes, we expect that constantly improving spatial and temporal resolution of experimental studies will reveal deviations from the Brownian dynamics model caused by memory effects, as already anticipated by simulations and theoretical studies\cite{straub_calculation_1987,straub_non-markovian_1986,berezhkovskii_single-molecule_2018,satija_broad_2020,satija_generalized_2019,medina_transition_2018,makarov_interplay_2013,grote_reactive_1981,grote_stable_1980,daldrop_butane_2018,ayaz_non-markovian_2021,dalton_fast_2023,lapolla_toolbox_2021,vollmar_model-free_2024}.   In this regard, we note that the division of the equilibrium ensemble into two nonequilibrium subensembles corresponding to reactants and products using splitting probabilities  (e.g., Eq. \ref{ground state decomposition}) remains valid even if the assumption of Langevin dynamics is no longer correct, although the calculation of the splitting probabilities is more challenging. 

\begin{acknowledgments}
\noindent This work was supported by the National Science Foundation (grant CHE 2400424 to DEM).  AMB is grateful to Eli Pollak, Attila Szabo, Peter Talkner, and Vladimir Zitserman  for numerous discussions of various aspects of the theory of activated rate processes. 
\end{acknowledgments}

\appendix

\section{Splitting probability for Langevin dynamics in a parabolic barrier}

The splitting probability $\phi((x,v)\to b|a)$ is a solution to the  Onsager-like equation\cite{onsager_initial_1938,banushkina_2016_nodate,risken_fokker-planck_1996} 
\begin{equation} \label{split PB from adjoint} 
    \hat{L}_{KK}^{*} \phi = 0,
\end{equation}
where 
\begin{equation} \label{LKKBck}
    \hat{L}_{KK}^{*}=v\frac{\partial}{\partial x}-\frac{U'(x)}{m}\frac{\partial}{\partial v}+D_v \frac{\partial}{\partial v} e^{\beta mv^2/2}\frac{\partial}{\partial v} e^{-\beta mv^2/2},
\end{equation}
is the backward  Klein-Kramers operator. It further must satisfy the absorbing boundary condition at $x=b$,
\begin{equation} \label{boundaryB)} 
    \lim_{x\to b-} \phi((x,v)\to b|a) = 1, \quad v>0,
\end{equation}
implying that the boundary $b$ will be crossed by  a particle situated next to the boundary and  moving toward it. 
To simplify the  notation, we  temporarily set
$$\phi((x,v)\to b|a) \equiv \phi(x,v)$$
throughout this Appendix. For the parabolic barrier, Eq.~\ref{parabolic barrier}, Eq.~\ref{split PB from adjoint} becomes
\begin{equation} \label{Onsager KK}
v \frac{\partial\phi(x,v)}{\partial x} + (\omega^2 x-\frac{\gamma v}{m}) \frac{\partial\phi(x,v)}{\partial v} +D_v \frac{\partial^2\phi(x,v)}{\partial v^2} =0
\end{equation}
Following Kramers\cite{kramers_brownian_1940}, we seek the solution in the form\footnote{the function $\phi(u)$ is the same as the function $\zeta(u)$ introduced by Kramers in his Eq. 19, and the following discussion closrely follows that of Kramers.}
\begin{equation} \label{Kramers ansatz}
    \phi(x,v)=\phi(u=v+\alpha x),
\end{equation}
where $\alpha$ is a constant to be determined later. This gives 
\begin{equation} \label{eq for phi}
\left[\omega^2x+(\alpha -\frac{\gamma}{m})v\right ]\frac{d\phi(u)}{du}+D_v \frac{d^2\phi(u)}{du^2}=0, 
\end{equation}
which implies that the expression in the square brackets must be proportional to $u=v+\alpha x$, or 
$$
\frac{\omega^2}{\alpha-\gamma/m} = \alpha .
$$
The parameter $\alpha$, therefore, satisfies a quadratic equation with the two roots
\begin{equation} \label{roots}
\alpha_{\pm} = \omega (\frac{\gamma}{2m\omega}\pm\sqrt{1+\frac{\gamma^2}{4m^2\omega^2}}).
\end{equation}
Let 
$$
g(u)=\frac{d\phi(u)}{du}.
$$
Then, solving Eq.~\ref{eq for phi} we find 
$$
g(u)=\zeta e^{-(\alpha_{\pm}-\frac{\gamma}{m})\frac{u^2}{2D_v}},
$$
where $\zeta$ is a constant to be determined later.  We now consider the case where the barrier between the boundaries $a$ and $b$ is much greater that $k_B T$. This allows us to replace, approximately,  the behavior of $\phi(u)$ near the boundaries by the behavior of $\phi(u)$ in the limits $u\to \pm \infty$. We then require that  
\begin{equation} \label{limits of phi 1}
\lim_{u\to -\infty}\phi(u)=0 
\end{equation}
and 
\begin{equation} \label{limits of phi 2}
\lim_{u\to \infty}\phi(u)=1. 
\end{equation}
The first of these two equations states that a particle that starts far to the left of the barrier top $x=0$  and/or moving fast to the left will never make it far to the right across the barrier. The second equation is an approximation to Eq.~\ref{boundaryB)}. 

Using Eq.~\ref{roots},  we now notice that only the root $\alpha_+$ makes physical sense, as the other one would produce an exponentially diverging solution at $|u|\to \infty$. Using this root, we have
\begin{equation} \label{g final}
    g(u)=\zeta e^{-\frac{\omega \kappa}{2D_v}u^2},
\end{equation}
where $\kappa$ is Kramers' transmission coefficient, Eq.~\ref{Kramers kappa}.  Finally, integrating this over $u$ and recalling  Eq.~\ref{limits of phi 2}, we obtain
\begin{equation}
    \phi(x,v)=\sqrt{\frac{\omega \kappa}{2\pi D_v}}\int_{-\infty}^{v+a_+ x} e^{-\frac{\omega \kappa}{2D_v}z^2} dz,
\end{equation}
which, using Eq.~\ref{roots}, can be seen to be equivalent to Eq.~\ref{splitting prob phase space}.

\section{More on the ansatz of Eq. \ref{ansatz}}
It is instructive to consider the function  $\hat{L}\tilde{\psi}_1 (x)$ obtained using the approximate Eq.~\ref{ansatz}. Using Eqs.~\ref{splitting} and \ref{integral I}, we have 
\begin{equation}
    \hat{L}\tilde{\psi}_1(x)=-(c_{A}+c_{B}) J_{a\to b}\left[ \delta(x-a)-\delta(x-b)\right],
\end{equation}
where $J_{a\to b}$  is the reactive flux defined by Eq. \ref{flux from I}.  Notably, we have 
\begin{equation}
    \hat{L}\tilde{\psi}_1(x)\ne \epsilon_1 \tilde{\psi}_1(x),
\end{equation}
as $\epsilon_1 \tilde{\psi}_1(x) \propto \tilde{\psi}_1(x)$ is a continuous function (Fig.\ref{fig:relaxation mode}) while  $\hat{L}\tilde{\psi}_1(x)$ turns out to be identically equal to zero everywhere except at $x=a,b$, where this function is singular!  In other words, while $\tilde{\psi}_1(x)$ is a reasonable approximation to $\psi_1(x)$, $\hat{L}\tilde{\psi}_1(x)$ does not seem to be a reasonable approximation to $\hat{L}\psi_1(x)$! Since the operator $\hat{L}$ is unbounded, this finding should not surprise us too much.  Quite remarkably, however, if we proceed to use $\tilde{\psi}_1$ instead of $\psi_1$ in Eq.~\ref{expected value eps1} to obtain a variational estimate of $\epsilon_1$, we find 
\begin{equation}
    (\tilde{\psi}_1,\hat{L}\tilde{\psi}_1)=-(c_{A}+c_{B}) J_{a\to b}\int_{-\infty}^{\infty}dx \left[ \delta(x-a)-\delta(x-b)\right] (c_{A} p_A(x)-c_{B} p_B(x))=-(c_{A}+c_{B})^2 J_{a\to b}
\end{equation}
Using $c_A$ and $c_B$ from Eq.\ref{lcao coefficients} and the fact that $P_A+P_B=1$, this can be rewritten as 
\begin{equation}
    (\tilde{\psi}_1,\hat{L}\tilde{\psi}_1)=\epsilon_1 =-(\frac{J_{a\to b}}{P_A}+\frac{J_{a\to b}}{P_B})=-(k_{A\to B}+k_{B\to A}),
\end{equation}
where $k_{A\to B}$ and $k_{B\to A}$ are the transition rates in the overdamped case,  as already established above, see Eq.~\ref{eps1 final}. This alternative derivation highlights that, despite failing to satisfy the eigenvalue equation Eq.~\ref{eig problem SM},  the approximate eigenfunction $\tilde{\psi}_1(x)$  gives an accurate variational estimate to the relaxation rate $-\epsilon_1$ for the double-well system, which is exactly equal to the prediction of the Kramers theory in the overdamped case. 

\section{Explicit expression for $\epsilon_1$ in terms of the splitting probabilities in the Langevin case and its relation to the equilibrium unidirectional fluxes.}
To evaluate the first eigenvalue $\epsilon_1$ in Eq.~\ref{Langevin eps1}, it is expedient to introduce the adjoint Klein-Kramers operator (which is obtained from the backward operator in Eq.~\ref{LKKBck} by changing the velocity sign, $v\to -v$), 
\begin{equation} \label{LKKAdj}
    \hat{L}_{KK}^{\dagger}=\hat{L}_{KK}^{*}(v\to-v)=-v\frac{\partial}{\partial x}+\frac{U'(x)}{m}\frac{\partial}{\partial v}+D_v \frac{\partial}{\partial v} e^{\beta mv^2/2}\frac{\partial}{\partial v} e^{-\beta mv^2/2},
\end{equation}
and its eigenfunctions
$$\chi_n(x,v)=\psi_n(x,v)/p_{eq}(x,v),$$
which have the same eigenvalues $\epsilon_{n}$, i.e., $\hat{L}_{KK}^{\dagger} \chi_n(x,v)=\epsilon_n \chi_n(x,v)$. 
The eigenfunctions $\psi_n$ and $\chi_n$ form a biorthogonal system,  obeying the equation

\begin{equation}
\begin{split}
 \int_{-\infty}^{\infty}dx \int_{-\infty}^{\infty}dv \chi_m(x,-v) \psi_n(x,v) \\=(\psi_n,\psi_m) \\
=\int_{-\infty}^{\infty}dx \int_{-\infty}^{\infty}dv \psi_m(x,-v) \psi_n(x,v)/p_{eq}(x,v) \\
= \int_{-\infty}^{\infty}dx \int_{-\infty}^{\infty}dv \chi_m(x,-v) \chi_n(x,v)p_{eq}(x,v) =\delta_{nm}
\end{split}
\end{equation}

Using  the ansatz of Eq.~\ref{ansatz phase space}, we have 
\begin{equation}
\tilde{\chi}_1(x,v)=\tilde{\psi}_1(x,v)/p_{eq}(x,v)=c_{A} \phi((x,-v)\to a|b)-c_{B}  \phi((x,-v)\to b|a),
\end{equation}
which is an (approximate) eigenfunction of the adjoint Klein-Kramers operator $\hat{L}^{\dagger}_{KK}$.  We then estimate the corresponding eigenvalue as

\begin{equation}
\epsilon_1 \approx  \int_{-\infty}^{\infty}dx \int_{-\infty}^{\infty}dv p_{eq}^{-1}(x,v)\psi_1(x,-v) L_{KK}\psi_1(x,v) = \int_{-\infty}^{\infty}dx \int_{-\infty}^{\infty}dv p_{eq}(x,v)\chi_1(x,-v) L^{\dagger}_{KK}\chi_1(x,v).
\end{equation}
Integrating this by parts, it can be shown that:
    \begin{multline} \label{eps 1 variational Langevin}
    \epsilon_1 \approx -\frac{2}{\beta m} \int_a^b dx \int_{-\infty}^{\infty}dv p_{eq}(x,v) \left[   \frac{\partial \tilde{\chi}_1(x,-v)}{\partial v} \frac{\partial \tilde{\chi}_1(x,v)}{\partial x} + \frac{\gamma}{2m} \frac{\partial \tilde{\chi}_1(x,-v)}{\partial v} \frac{\partial \tilde{\chi}_1(x,v)}{\partial v}  \right]\\
    =-\frac{2}{\beta m P_A P_B} \int_a^b dx \int_{-\infty}^{\infty}dv p_{eq}(x,v) \frac{\partial \phi((x,v)\to b|a)}{\partial v} \left[ \frac{\partial}{\partial x}+ \frac{\gamma}{2 m} \frac{\partial}{\partial v}\right] \phi((x,-v)\to b|a), 
    \end{multline}
    which is the phase-space analog of Eq.~\ref{eps1 approx}.   Eq.~\ref{Langevin eps1} further implies that the equilibrium unidirectional flux must be given by
 \begin{equation} \label{flux from eps1 Langevin}
     J_{a\to b}=\frac{2}{\beta m} \int_a^b dx \int_{-\infty}^{\infty}dv p_{eq}(x,v) \frac{\partial \phi((x,v)\to b|a)}{\partial v} \left[ \frac{\partial}{\partial x}+ \frac{\gamma}{2 m} \frac{\partial}{\partial v}\right] \phi((x,-v)\to b|a)
 \end{equation}    
 Using  the approximate splitting probability for a harmonic barrier, Eq.~\ref{splitting prob phase space}, one can show that this result leads to Eq.~\ref{a to b flux moderate friction}.  Therefore, for a sufficiently high harmonic barrier, the method of estimating the reactive flux based on the first eigenvalue of the Klein-Kramers equation is equivalent to the exact calculation of the reactive flux between the boundaries $a$ and $b$ given by Eq.~\ref{a to b flux phase mod}. Because an exact analytical solution for the phase-space splitting probability is unknown, we have not been able to verify whether or not Eqs.~\ref{a to b flux phase mod} and  \ref{flux from eps1 Langevin} are formally equivalent similarly to the diffusive case, where -- somewhat unexpectedly -- the variational calculation yields a result that is identical to the exact flux of Eq.~\ref{uni flux diff} (see Eq.~\ref{eps1 final}).

\bibliography{references.bib}
\end{document}